\documentclass[acus]{JAC2000}

\newcommand{\mytt}[1]{\texttt{\small{#1}}}

\begin{document}

\title{\flushright{WEAP034}\\[15pt] 
    \centering CONVERTING EQUIPMENT CONTROL SOFTWARE \\ FROM PASCAL TO C/C++}
\author{L. Hechler, GSI, Darmstadt, Germany}

\maketitle

\begin{abstract}
The equipment control (EC) software of the GSI accelerators has 
been written entirely in Pascal. Modern software development 
is based on C++ or Java. To be prepared for the future, we 
decided to convert the EC software from Pascal to C in a first 
step. Considering the large amount of software, this is done 
automatically as far as possible. The paper describes our 
experiences gained using a Pascal to C translator, Perl 
scripts, and, of course, some manual interaction.
\end{abstract}

\section{MOTIVATION}
The EC software comprises the device representation layer, 
the real-time layer, and the device drivers \cite{siscos}. 
Except for some assembler code, it has been written 
entirely in Pascal. 

For embedded applications there are no integrated cross 
development systems that support Pascal any more. The 
system we use runs under VMS and its support expires 
completely by the end of 2001.

However until now we invested about 40 person-years 
in developing and maintaining EC software. A lot 
of special know-how has gone especially into the 
real-time layer. The functionality gained in this 
work must be preserved.

Future control system developments have to be realized
with modern object-oriented methodes. Appropriate
up-to-date tools are based on C++ or Java nearly
without exception.   

Existing hardware (400 VME boards) has to be used 
in the future as well since it cannot be replaced 
completely due to cost reasons. And, last but not least, 
the conversion must not affect the day by day accelerator 
operation.

\section{CONVERSION}

We decided to convert the EC software from Pascal to C 
in a first step. This allows us to ``re-use'' the software
on one hand and to establish a basis for re-engineering 
the control system with modern methods and tools \cite{reeng}
on the other hand. 

Considering that EC software consists of about 170\,000 
lines of code (LOC), comments not counted, it is clear that 
conversion has to be done automatically as far as possible.

The basis for a conversion is EC software for 
one device class. There are 61 different device classes, 
each one controlled by dedicated software. To ease the 
conversion, we issued a cookbook \cite{cookbook} that 
describes the process step by step.

To convert the Pascal code into C automatically we use
the Pascal to C translator p2c\footnote{p2c is part of many
    Linux distributions. It runs under VMS as well.}.

Perl scripts are used to adapt the notation of identifiers
to our style guide.

In spite of the automation 
there is a lot of manual interaction left over.
Beside the preparations for p2c and Perl there are four
essential reasons that make manual interaction neccessary.

\subsection{Compatibility of Data Structures}

The p2c manual pages state that ``most reasonable Pascal 
programs are converted into fully functional C which will 
compile and run with no further modifications''. This may
be true for stand-alone programs. Given the EC software it
has to be taken into account that in case of communications
with other modules, e.\,g.\ programs of the operating layer,
the structure of interchanged data has to be kept fully 
compatible, because those modules have not been changed.

1.  Pascal supports \mytt{PACKED} records und arrays 
    to facilitate minimal alignment space between
    elements. C does not support this feature.

2.  At GSI we use the Organon Pascal compiler from 
    CAD-UL which supports the dialect of the Oregon Pascal/2 compiler. 
    Their syntax only differ in one key word, but they generate
    completely different codes. However p2c makes some assumptions about 
    the generated code, e.\,g.\ the order of bits in a bitset,
    which is crucial for instance when hardware registers are accessed.

3.  In Pascal the allocation size of an enumeration type 
    depends on the number of its elements. It may be one or two bytes. 
    In C the allocation size is always an \mytt{int}.

4.  The Pascal string \mytt{ARRAY [1..len] OF CHAR} contains
    \mytt{len} characters. Its allocation size in memory is 
    \mytt{len} bytes. A C string with equal size is 
    \mytt{char s[len]}. It can hold only \mytt{len} - 1 
    (printable) characters because of the terminating 
    {\small\verb|\0|}.

\subsection{Linking of Pascal and C modules}

A CPU of the device representation layer hosts EC software
of up to 12 different device classes. On this layer it must
be possible to combine modules written both in Pascal and C
because EC software for a number of device classes can not be 
converted at the same time.

Combining Pascal and C modules means that they have to be
linked together. In this case identical procedure calling 
mechanisms have to be ensured.

1.  P2c translates routine parameters into a structure that contains
    a pure C function pointer and a ``static link'', a pointer to 
    the parent procedure's locals. This structure is passed to the 
    called function. Both of our compilers, the Pascal as well as
    the C compiler, need plain C function pointers. The
    option to force p2c to use this concept is available but it
    does not work.

2.  Pascal can handle conformant array routine
    parameters defined as
{\small
\begin{verbatim}
    f(a: ARRAY [lo..hi: INTEGER] OF MyType);
\end{verbatim}
}
    by syntactically passing the array as actual parameter only:
{\small
\begin{verbatim}
    VAR x: ARRAY [7..13] OF MyType;
    f(x);
\end{verbatim}
}
    On calling the routine, the array, or its address in case of 
    a \mytt{VAR} parameter, as well as the lower and upper limit of 
    the array are pushed onto the stack. Thus the array bounds may
    be checked by the called routine. 

    P2c generates C code where the routine is declared and 
    called with three parameters explicitly. The order the
    parameters are pushed onto the stack differs from that
    of the Pascal compiler.

\subsection{Maintainability}
The converted software is no final product. It has to be 
maintained for changed or extended future requirements. 
Therefore readable and comprehensible code is indispensable.
To achieve this sufficient work has to be invested into
simplifying and refurbishing the plain C code produced by
p2c and the Perl scripts\footnote{Due to restricted space
    only some items are mentioned here.}.

1.  Pascal supports nesting of routines. The parent 
    routine's local variables lie in the scope of the nested 
    routine. C does not provide this concept. So p2c combines 
    the parent routine's locals to a single structure and adds 
    an additional link parameter to the sub-routine's parameter 
    list that points to this structure thus allowing the 
    sub-routine to access its parents' variables. C code 
    designed like this looks somewhat odd.

2.  Pascal provides the \mytt{WITH} statement to 
    abbreviate the notation for references to fields of structured 
    variables.
{\small
\begin{verbatim}
    WITH struc.field DO subfield := 1;
\end{verbatim}
}
    P2c creates a pointer for every \mytt{WITH} statement with
    generated names \mytt{WITH}, \mytt{WITH1}, \mytt{WITH2}, 
    etc.\ to access the field of a structure.
{\small
\begin{verbatim}
    T_field *WITH = &struc.field;
    WITH->subfield = 1;
\end{verbatim}
}
    Often there is no explicit type for the field the \mytt{WITH}
    statement references. In those cases p2c needs to declare an 
    additional pointer type first (\mytt{typedef struct T\_field ...}) 
    before it can define the pointer itself. 

    These constructs are hardly found in common C programs as well.

3.  Pascal allows to define an array of structures within
    one statement. A variable definition looks like this:
{\small
\begin{verbatim}
    VAR x: ARRAY [1..7] OF 
             RECORD i: INTEGER; c: CHAR END;
\end{verbatim}
}
    Although C supports a corresponding construct,
    p2c declares a structured type before it defines the 
    array.
{\small
\begin{verbatim}
    typedef struct _REC_x {int i; char c} _REC_x;
    _REC_x x[7];
\end{verbatim}
}
    To do so p2c must generate a name for the structured type, 
    which is \mytt{\_REC\_x} where \mytt{x} is the name of the 
    array.

\subsection{P2C Errors}

We encountered only two substantial p2c errors not mentioned
in the p2c manual. Both of them are very difficult
to detect since the compiler does not report an error. 
Overlooking them during the manual interaction means
they occur during the runtime of the software where they
are moreover hard to debug.

1.  In some cases p2c translates a Pascal 32 bit wide 
    unsigned integer type 
{\small
\begin{verbatim}
    TYPE uns_long = 0..16#FFFFFFFF;
         myType = uns_long;
\end{verbatim}
}
    into a single C character type.
{\small
\begin{verbatim}
    typedef char myType;
\end{verbatim}
}
    The error occurs only unfrequently. Unfortunately we were
    not able to reproduce the circumstances of its occurrence.

2.  The Pascal pointer \mytt{ptr} should point to a 16 bit wide 
    type, e.\,g.\ a hardware register, that has an offset 
    of 4 bytes to a base address \mytt{addr}.
{\small
\begin{verbatim}
    TYPE uw_p = ^uns_word;
    VAR  ptr: uw_p;
         addr: uns_long;
    ptr := loophole(uw_p, addr + 4);
\end{verbatim}
}
    In rare cases p2c translates the pointer assignment to
{\small
\begin{verbatim}
    ptr = (uns_word*)((uns_long*)addr + 4);
\end{verbatim}
}
    which results in a miscalculated pointer value.
    The expression \mytt{(uns\_long*)addr} type-casts
    \mytt{addr} to a pointer to a 32 bit type and thus
    adding 4$\times$4 = 16 bytes to the base address
    instead of 4.

\section{APPLYING THE STYLE GUIDE}
Unlike C Pascal identifiers are case insensitive. P2c takes
the first occurence of an identifier to determine the notation
of all subsequent occurences. Mostly these notations do not 
conform to our style guide. To force the notation of 
identifiers according to the style guide, we developed some 
Perl scripts that do most of the job. 

A Perl script recognizes expressions for instance like
{\small
\begin{verbatim}
    #define The_Answer 42
    typedef struct my_type {...} my_type;
\end{verbatim}
}
and suits the identifiers accordingly (getting
\mytt{THE\_ANSWER} and \mytt{MyType}).

To handle more complex constructs a parser-like 
script would be required. This is not implemented 
yet. Thus manual modifications are neccessary 
whereby each identifier has to be adjusted only 
once.

All changes of identifiers in the software of one 
device class are then stored as key value pairs in 
a device class specific local data base (DB). The
pairs describe the translation from the old into
the new style guide conform notation. The creation 
and completion of the local DB is done by another
Perl script.

A third script is used to apply the translations 
stored in the local DB to all identifiers in all 
files of a device class. Additionally a global
DB is used which applies the translation of the 
identifiers of the system interface.

\section{STATUS}

Meanwhile EC software for 15 different device classes 
has been converted. Devices are operated with the 
converted software since more than 6 months. Some of 
them even in therapy operation \cite{therapy}.
Apart from teething troubles in the beginning of the conversion 
process the software has showed good quality and bug fixing is 
an amazingly rare necessity.

\subsection{Time}

To estimate the manual interaction effort to convert the 
software for one device class the process can be split 
into 4 phases. The outcome is the following distribution:

\begin{tabular}{lll}
1. & p2c including some preparations & 10\% \\
2. & manual interaction, part I & 20\% \\
3. & Perl including building of local DB & 20\% \\
4. & manual interaction, tests, bug fixing & 50\% \\
\end{tabular}

Phase 2 is neccessary since some manual interactions are 
better done before using the Perl scripts. Although 1 and 3
are the ``automatic'' phases they also need manual actions,
particularly phase 3. With more experience the percentage
of phase 4 increases but the overall conversion time 
decreases.\smallskip

On an average EC software for one device class consist of
2200 LOC. Its conversion requires us about 2 person-weeks.
To convert the whole EC software consisting of 170\,000
LOC we will need approximatly 39 person-months or 
3.25 person-years.

Without the help of p2c and Perl scripts we roughly
estimate twice to four times the effort. There was
only one attempt to convert a device class completely 
manually.

Balzert \cite{swtech} states that software \emph{development}
results in 350 LOC per person-month.

Given this our method is 2 to 4 times faster than a pure
manual conversion and more than 10 times faster than a
redevelopment.

\section{CONCLUSION}
Using p2c and Perl scripts converting EC software from 
Pascal to C is feasible without major problems. In spite
of the automation tools there is a lot of manual interaction 
left over until C software for a device class is ready 
to be released.

Our method allows us to convert EC software in
reasonable time. Entirely re-engineering the EC software
would have exceeded our manpower capacity excessively.      

With EC software converted to C we are well-prepared
to take the next step to C++ (or Java). It should be 
possible at least on the device representation layer 
to re-use the C functions, which are usually straight
forward, as methods of classes in C++. The use of C++ on
the real-time layer has to be investigated, particularly
with regard to the highly demanding 50\,Hz linear
accelerator operation.

\section{ACKNOWLEDGEMENTS}
Thanks to Peter Kainberger for all the Perl scripts,
and to him, Gudrun Schwarz, and Regine Pfeil for 
contributions to the cookbook.


\begin{thebibliography}{9}

\bibitem{siscos}
U. Krause, V. Schaa, R. Steiner, ``The GSI Control System'', Proceedings
of ICALEPCS '91, Tsukuba, Japan, 1991.

\bibitem{reeng}
U. Krause, ``Re-Engineering of the GSI Control System'', these proceedings.

\bibitem{cookbook}
L. Hechler, P. Kainberger, G. Schwarz, ``P2C - Umstellung der 
Ger\"atesoftware von Pascal nach C/C++'', Accelerator Controls 
Documentation U-GSW-08, GSI, Darmstadt, November 2000,
http://bel.gsi.de/mk/sty/p2c.html.

\bibitem{therapy}
U. Krause, R. Steiner, ``Adaption of a Synchrotron Control System
    for Heavy Ion Tumor Therapy'', Proceedings of ICALEPCS '95, 
    Chicago IL, USA, 1995.

\bibitem{swtech}
Helmut Balzert, ``Lehrbuch der Software-Technik: Software-Entwicklung'', 
Spektrum Akad.\ Verlag GmbH, Heidelberg, Berlin, Oxford, 1996.

\bibitem{p2c}
Dave Gillespie, ``p2c - Pascal to C Translator Manual Pages'', Caltech.

\bibitem{cstyle}
Udo Krause, ``C/C++ Style Guide'', Accelerator Controls Documentation 
O-SIS-10, GSI, Darmstadt, December 2000,
http://bel.gsi.de/mk/sty/cstyle.html.

\end{thebibliography}
\end{document}